\documentstyle[aps,prl,epsfig,multicol,graphicx]{revtex}

\begin{document}
\draft

\wideabs{
\title{Communication in networks with hierarchical branching}
\author{A. Arenas$^1$, A. D\'{\i}az-Guilera$^2$, and R. Guimer\`a$^3$}
\address{$^1$Departament d'Enginyeria Inform\`atica, Universitat
Rovira i Virgili, Ctra. Salou s/n, 43006 Tarragona, Spain\\
$^2$Departament de F\'{\i}sica Fonamental, Universitat de Barcelona,
Diagonal 647, 08028 Barcelona, Spain\\
$^3$Departament d'Enginyeria Qu\'{\i}mica, Universitat Rovira i Virgili,
Ctra. Salou s/n, 43006 Tarragona, Spain}

\date{\today}
\maketitle

\begin{abstract}
We present a simple model of communication in networks with hierarchical branching. We analyze the behavior of the model from the viewpoint of critical systems under different situations. For certain values of the parameters, a continuous phase transition between a sparse and a congested regime is observed and accurately described by an order parameter and the power spectra. At the critical point the behavior of the model is totally independent of the number of hierarchical levels. Also scaling properties are observed when the size of the system varies. The presence of noise in the communication is shown to break the transition. Despite the simplicity of the model, the analytical results are a useful guide to forecast the main features of real networks.
\end{abstract}

}

Nowadays, many challenging questions have arisen concerning the behavior
of complex technological, economical, and social systems
\cite{arthur97}. In particular, computer simulations of agents and their
interactions (agent-based modeling) has become a widely used tool in our
current understanding of their macroscopic behavior \cite{axelrod97}.
Specially interesting is the study of
hierarchical branching in networks because it seems to be the basic structure underlying
complex organizational systems. Our interest is focused on the behavior of
hierarchical structures formed by elements (or agents) that interact
with each other via communication processes. This framework is
especially adequate to study e.g. Internet flow
\cite{csabai94,tretyakov98,takayasu96,ohira98,sole??}, 
traffic networks \cite{chowdhury00}, river networks 
\cite{banavar99} and even communication flows in organizations 
\cite{radner93}.

In this letter, we propose and study a very simple model of
communication. The model includes only the
basic ingredients present in a communication process between two
elements: (i) information packets to be transmitted (delivered) and (ii)
communication channels to transmit the packets.
Despite its simplicity, the model reproduces the main characteristics of
the flow of information packets in a network, and is general enough to allow the study of communication processes in many conditions: for example different capabilities of agents to transmit packets, and/or heterogeneity in the communication channels (miscommunication, exogenous effects, etc.) represented by introducing disorder. We observe three different behaviors depending on the capability of agents to transmit packets. In particular, for a certain capability, we observe a continuous phase transition between a sparse and a congested regime when the number of packets to deliver reaches a critical value. Near the transition point signs of criticality arise, we find large fluctuations, critical slowing down and power law behavior of power spectrum of the amount of information flowing in the network, in agreement with reported empirical data \cite{takayasu96}. We provide a mean field estimation of the critical point in good agreement with simulation results and we define analytically an order parameter to characterize the behavior of the system.

\begin{figure}[h]
\centerline{\includegraphics*[width=0.8\columnwidth]{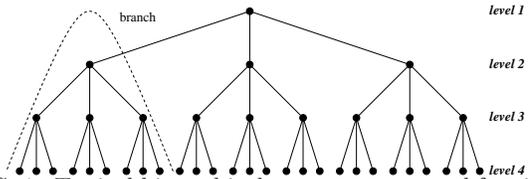}}
\caption{Typical hierarchical tree structure used for simulations and
calculations: in particular, it is a tree $(3,4)$. Dashed line:
definition of {\it branch}.
\label{network}}
\end{figure}

The model is defined in the following way: the communication network is
mapped onto a lattice where nodes represent the communicating elements
(for instance, employees in a company, routers and servers in a computer
network, etc.) and the links between them represent communication lines. In
particular, we use hierarchical trees as depicted if Fig.~\ref{network}, although most of the results reported hold when considering that the hierarchical branching is characteristic of the paths that information follows and not of the topology of the network itself.
These structures are characterized by two quantities: the branching
factor, $z$, and the number of levels, $m$. From now on, we will use the
notation $(z,m)$ to describe a particular tree.


The dynamics of the model is the following. At each time step $t$, an
information packet is created at every node with probability $p$. When a
new packet is created, a destination node, different from the origin
node, is chosen at random in the network. Thus, during the following
time steps $t, t+1,\ldots ,t+T$, the packet is traveling towards its
destination: once the packet reaches this destination node, it is
delivered (disappears from the network). The time a packet remains in
the network is related not only to the distance between the source node
and the target node, but also to the amount of packets in the network.
In particular, at each time step, all the packets move from their
current position, $i$, to the next node in their path, $j$, with a
probability $q_{ij}$. We define $q_{ij}$, {\it quality of communication}
between $i$ and $j$, as
\begin{equation}
q_{ij}=\sqrt{k_{ij}k_{ji}}.
\end{equation}
where $k_{\alpha\beta}$ represents the capability of agent $\alpha$ to
communicate with agent $\beta$ at each time step. For $k_{\alpha\beta}$
we propose:
\begin{equation}
k_{\alpha\beta}=\xi_{\alpha\beta}f(n_\alpha)
\end{equation}
where $\xi_{\alpha\beta}$ is a uniformly distributed random number in
the interval $[0,1]$ representing the effects mentioned above for the
directional connection between $\alpha$ and $\beta$
\cite{communication}, $n_\alpha$ is the total number of packets
currently at node $\alpha$, and $f(n)$ determines how the capability evolves when the number of packets at a given node increases.

Since any election of $f(n)$ could be valid, we
will study the general form
\begin{equation}
f(n)=\left\{ \begin{array}{lcl}
        1		&       \quad\mbox{for} &       n=0\\
        n^{-\gamma}	&       \quad\mbox{for} &       n=1,2,3,\ldots
        \end{array}\right.
\label{f_1}
\end{equation}
with $\gamma\geq0$. The average number of packets delivered by a node $\alpha$ to another node $\beta$ will be proportional to $n_\alpha/(n_\alpha^{\gamma/2}n_\beta^{\gamma/2})$. Assuming the degree of homogeneity derived from the model, $n_\alpha\sim n_\beta$, the former expression reads $n_\alpha^{1-\gamma}$. It is straight forward to recognize three different behaviors corresponding to three different values of $\gamma$ in the previous formula. For $\gamma>1$, the number of transmitted packets decreases as $n_\alpha$ grows. For small values of the probability of packet generation per node and time step, $p$,  all the packets are delivered and hence, after a transient, the system reaches a steady state in which the total number of packets, $N$, fluctuates around a constant value. However, if we increase $p$ at some point the total number of packets will be so large that the network will not be able to handle them, $N$ will increase continuously and, at the end, no packets at all will be delivered to their destination. On the contrary, for $\gamma<1$, the number of transmitted packets grows as $n_\alpha$ does. Thus, the number of delivered packets increases as $N$ grows until an equilibrium between generated and delivered packets is reached: at this point, $N$ remains constant (except fluctuations). In case $\gamma=1$, the number of delivered packets is constant irrespective of the number of stored packets (note that this is consistent with simple models of queues \cite{ohira98}). This particular behavior is less obvious and will be treated accurately from the viewpoint of critical systems.

\begin{figure}[h]
\centerline{\includegraphics*[width=0.8\columnwidth]{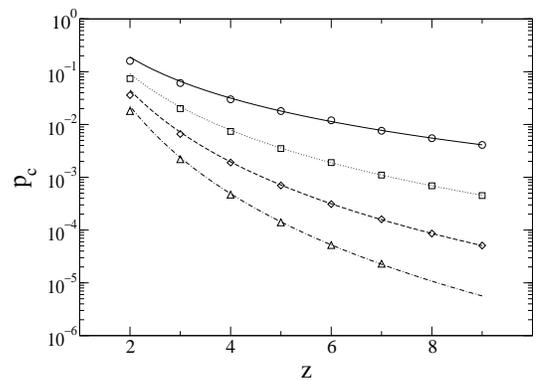}}
\caption{Comparison between simulated (symbols) and analytical (lines)
values for the critical probability of packet generation, $p_c$ as a
function of the branching factor $z$ for hierarchies with different
number of levels: $m=4$ (circles and full line), $m=5$ (squares and
dotted line), $m=6$ (diamonds and dashed line) and $m=7$ (triangles and
dot-dashed line). The error bars are smaller than the symbol size.
\label{p_c}}
\end{figure}

As a first step, let us concentrate in the case $\xi_{ij}=1\mbox{,
}\forall i,j$. From simulations, we observe two different regimes and, as in the case $\gamma>1$, $p$ plays the role of a control parameter. For small values of $p$, all the packets are delivered while for
large values of $p$, not all the packets can reach their destination, 
and $N$ grows in time with no limit. The key point is that, since the number of delivered packets is independent of $N$, there is always a fraction of problems that are reaching their destination and the transition to the collapsed regime is continuous. This transition
occurs for a critical value of $p$, $p_c$, whose exact value depends on
the network parameters $z$ and $m$ (see Fig.~\ref{p_c}). For values of
$p$ smaller than but close to $p_c$, the steady state is reached but
large fluctuations with long correlation times appear.

At the subcritical region, the power spectrum of the total number of
packets, $N(t)$, is well fitted by a Lorentzian characterized by a
certain frequency, $f_c$. As we get closer to $p_c$, we observe that
$f_c\rightarrow0$ and the power spectrum becomes $1/f^2$ for the whole
range of frequencies. That means that the  average time the packets
remain in the network grows as we approximate the critical point
(critical slowing down). We have also analyzed the power spectrum of the
number of packets at individual nodes, $n_i(t)$. The main result is that
the power spectrum of $N(t)$ is dominated by the top node which is the
most congested: near $p_c$, the power spectrum for this node is also
$1/f^2$. As one goes down in the hierarchy the number of packets
diminishes and the power spectra have $1/f^\beta$ tails with $\beta$
decreasing from 2 to 0 at the lowest level. The last result is
consistent with the fact that the bottom agents deliver packets
immediately and so $n_i(t)$ is a time series of peaks separated by
Poisson distributed time intervals. As it is well known, this kind of
series have white noise spectra. We have also checked other topologies
\cite{guimera??} and found that in a square lattice with closed
boundaries the central sites have $\beta\sim1.2$ (in agreement with
Refs.~\cite{sole??,takayasu96}) whereas agents close to the boundaries
are less congested and a much lower exponent for the tail ($\beta\sim
0$) is observed.

As happens in other problems in statistical physics \cite{stauffer92},
the particular symmetry of the hierarchical tree allows a mean field
estimation of the critical point $p_c$ (although these calculation can
be performed under more general conditions \cite{guimera??}). Since in
the steady state regime there is no accumulation of packets, the number
of packets arriving to the top of the hierarchical structure (level 1)
per time unit, $n_1^a$, is, on average, equal to the number of packets
that are created in one branch of the network and have their destination
in a different branch (see Fig.~\ref{network}). Since the origin and the
destination of the packets are chosen at random, from purely geometric
considerations it is straightforward to estimate this number of packets
per unit time as:
\begin{equation}
n_1^a=p\left(\frac{z\left(z^{m-1}-1\right)^2}{z^m-1}+1\right).
\label{n_1^a}
\end{equation}
Within this mean field approach, it can be easily shown that it is
indeed the top node which is the most congested.

On the other hand, in our mean field calculation $q_{12}$ is the average
probability that a given packet moves from a node in the second level to
the top node and vice versa, and is given, as a first approximation, by
$q_{12}=1/\sqrt{n_1n_2}$, where $n_1$ is the average number of packets
at level one and $n_2$ is the average number of packets at each of the
$z$ nodes in the second level. Thus the average number of packets
leaving the top at each time step will be $n_1^l=n_1q_{12}$, and the
average number of packets going from the $z$ nodes in the second level
to the top will be $n_1^a=z\alpha n_2q_{12}$, where $\alpha$ stands for
the fraction of packets in the second level that are trying to go up
(some of the packets in level 2 are, of course, trying to go down to
level 3).

At the critical point the top agent becomes collapsed and the
communications between the first and the second level are much more
congested than the communications between levels 2 and 3 so we can
assume that $\alpha\approx 1$. At this point, by imposing the steady
state condition $n_1^a=n_1^l$ we arrive to the relations $n_1=zn_2$ and
$n_1^a=\sqrt{z}$. Using equation (\ref{n_1^a}) we obtain the final
expression for $p_c$:
\begin{equation}
p_c=\frac{\sqrt{z}}{\frac{z(z^{m-1}-1)^2}{z^m-1}+1}
\label{p_c_value}
\end{equation}

Although strictly speaking the condition $\alpha=1$ provides an upper
bound to $p_c$, equation (\ref{p_c_value}) is an excellent approximation
for $z\ge 3$, as depicted in Fig.~\ref{p_c}.


The critical total number of generated packets, $N_c=p_cS$, with $S$
standing for the size of the system, can be approximated, for large
enough values of $z$ and $m$ such that $z^{m-1}\gg1$, by
\begin{equation}
N_c=\frac{z^{3/2}}{z-1},
\end{equation} 
which is independent of the number of levels in the tree. It suggests
that the behavior of the top node is only affected by the total number
of packets arriving from each node of the second level, which is
consistent with the mean field hypothesis.

\begin{figure}[h]
\centerline{\includegraphics*[width=0.8\columnwidth]{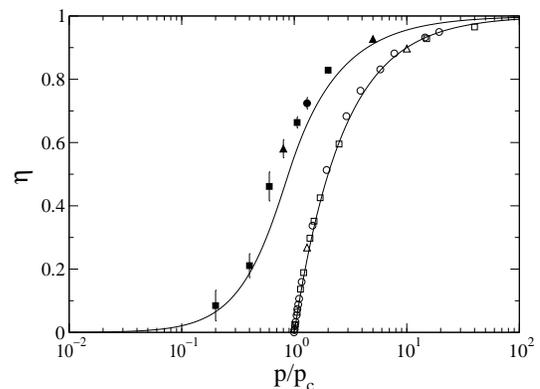}}
\caption{Behavior of the order parameter in both cases with noise
(filled symbols) and without noise (open symbols), for different
structures: $(6,7)$ (circles), $(3,6)$ (squares), $(5,4)$ (triangles).
The lines represent analytical results obtained for the simplest case
$(1,2)$.
\label{order}}
\end{figure}

In order to characterize the transition, we introduce an order
parameter:
\begin{equation}
\eta(p)=\lim_{t\rightarrow\infty}\frac{1}{pS}\frac{\left\langle\Delta N\right\rangle}{\Delta t},
\label{def_order}
\end{equation}
where $\Delta N=N(t+\Delta t)-N(t)$ and $\langle\ldots\rangle$ indicates
average over time windows of width $\Delta t$. Essentially, this order
parameter represents the ratio between undelivered and generated
packets at the stationary state. For $p>p_c$, the system collapses, $\langle\Delta N\rangle$
grows linearly with $\Delta t$ and thus $\eta$ is a function of $p$
only. For $p<p_c$, $\langle\Delta N\rangle=0$  and $\eta=0$. As observed
in Fig.~\ref{order}, and as may be expected from a properly chosen order
parameter, when $p$ is rescaled with $p_c$, the form of $\eta$ does not
depend on the details of the structure of the network, $z$ and $m$.


As far as $\eta$ does not depend on the structure of the network, we can
study the simplest case $(1,2)$ in order to obtain an analytical
estimation of the order parameter. In this case, the network consist of
only 2 nodes, 1 and 2, interchanging packets. Since from symmetry
considerations $n_1=n_2$, the maximum average number of delivered
packets per time unit will be $(n_1+n_2)/\sqrt{n_1n_2}=2$. Thus $p_c=1$
and with the present formulation of the model it is not possible to
achieve the supercritical regime. However, it is possible to extend $p$
to be the average number of generated packets per node and time step and
then $p$ can be greater than one. In this case, for $p>p_c$ the number
of packets delivered per time unit will be $2$ while the number of
generated packets will be $2p$. Thus
\begin{equation}
\eta=\frac{p-1}{p},
\end{equation}
in good agreement with simulated values (Fig.~\ref{order}). In
particular, near $p_c$ we have:
\begin{equation}
\eta\sim(p-p_c).
\end{equation}

Now let us consider the case where $\xi_{ij}$ take values uniformly
distributed in $[0,1]$. Even for very small values of $p$, a particular
realization of the disorder can provoke a very weak communication line
and the congestion of the whole network. Thus there is no transition
controlled by $p$. However, it is still possible to define the order
parameter as in (\ref{def_order}), just considering that the average
$\langle\ldots\rangle$ has to be taken over time and over disorder
realizations. As observed in Fig.~\ref{order}, the existence of disorder
destroys the phase transition acting as a random local magnetic field in
a paramagnetic-ferromagnetic transition \cite{villain84} and other
physical systems \cite{arenas93}.

Again, it is possible to obtain an analytical expression of the order
parameter in the case of two nodes. As in the ordered case, the number
of packets generated in a time step will be $2p$. Now, however, for a
particular realization, $\xi_{12}\mbox{ and }\xi_{21}$, the maximum
number of delivered packets will be $2\sqrt{\xi_{12}\xi_{21}}$. Thus, if
$\xi_{12}\xi_{21}>p^2$ the system will reach the steady state and the
configuration will not contribute to the order parameter, while if
$\xi_{12}\xi_{21}<p^2$ the system will collapse and the contribution
will be $\eta_{\xi_{12}\xi_{21}}=1-\sqrt{\xi_{12}\xi_{21}}/p$.

Thus we can define:
\begin{equation}
\eta(p,\xi_{12},\xi_{21})=\left\{ \begin{array}{lcl}
        0 & \quad\mbox{for} & \xi_{12}\xi_{21}>p^2\\
        1-\sqrt{\xi_{12}\xi_{21}}/p & \quad\mbox{for} & \xi_{12}\xi_{21}<p^2
        \end{array}\right.
\end{equation}
and the order parameter will be given by the average over the random
variables:

\begin{equation}
\eta(p)=\int\limits_0^1d\xi_{12}\int\limits_0^1d\xi_{21}\eta(p,\xi_{12},\xi_{21}).
\end{equation}

It is straightforward to obtain the result:
\begin{equation}
\eta(p)=\left\{ \begin{array}{lcl}
        1-4/(9p) & \quad\mbox{for} & p>1\\
        \left(5p^2-3p^2\ln{p^2}\right)/9 & \quad\mbox{for} & p<1
        \end{array}\right.
\end{equation}

As depicted in Fig.~\ref{order}, there is reasonable agreement between
this analytical expression and the points obtained by simulation, always
keeping in mind the simplicity of our approach.

Summarizing, we have studied a simple and general model of communication
in a network with hierarchical branching. We have obtained some
analytical results defining an order parameter and studying its behavior with
respect to the relevant parameters of the model. The behavior of the system at the critical regime shows to be independent of the number of levels in the hierarchy. This phenomena shows that the main features of information flow in a network with hierarchical branching is determined by the branching of the first level. Although we are in a very temptative stage of the model, we think that this result can help to understand flow in real networks, where this effect can dominate the global behavior of the system. Another interesting issue is the scaling observed in Fig.~\ref{order}. From the viewpoint of organizational design, this scaling can be used to forecast the behavior of the organization when increasing or decreasing its size. The inclusion of a
quenched randomness accounting for different kinds of interaction is not
a hindrance for our theoretical analysis and we give an accurate
behavior of the order parameter in this situation. The approach
presented here opens a line of research which will follow to
study different dynamics and topologies. 

\acknowledgments
The authors are gratefully acknowledged to F. Giralt, C.J. P\'erez, F. Vega and H.J. Witt
for helpful comments. This work has been supported by DGES of the
Spanish Government, grants PB96-0168, PB96-1011 and PB96-1025, and EU
TMR grant ERBFMRXCT980183. R.G. also acknowledges financial support from
the Generalitat de Catalunya.

\begin{references}
\bibitem{communication}We consider
$\xi_{\alpha\beta}\neq\xi_{\beta\alpha}$. Conceptually, our election is
more adequate for systems where agents determine the efficiency of the
communication process (as in organizations) while the symmetric case
$\xi_{\alpha\beta}=\xi_{\beta\alpha}$ would be more adequate for systems
where the channel does so (as in computer networks traffic). Although
the particular results obtained would be different in this case, the
qualitative behavior of the system would be the same and similar
calculations should be performed.
\end{references}
\end{document}